# Size-Dependent Luminescence Properties of Chromatographically-Separated Graphene Quantum Dots

*Naoto Fuyuno,[†] Daichi Kozawa,[†] Yuhei Miyauchi,[†,‡] Shinichiro Mouri,[†] Ryo Kitaura,[§] Hisanori Shinohara,[§] Toku Yasuda,[∥,⊥] Naoki Komatsu,[∥] and Kazunari Matsuda\*[†]*

[†]Institute of Advanced Energy, Kyoto University, Uji, Kyoto 611-0011, Japan

[‡]Japan Science and Technology Agency, PRESTO, 4-1-8 Honcho Kawaguchi, Saitama 332-0012, Japan

[§]Department of Chemistry and Institute for Advanced Research, Nagoya University, Nagoya 464-8602, Japan

[∥]Department of Chemistry, Shiga University of Medical Science, Otsu 520-2192, Japan

[⊥]Department of Materials Science, Osaka Electro-Communication University, Neyagawa, Osaka 572-8530, Japan

**Corresponding Author**

\*E-mail address: matsuda@iae.kyoto-u.ac.jp.

## ABSTRACT

We studied the photoluminescence (PL) properties of graphene quantum dots (GQDs) separated by size-exclusion high performance liquid chromatography. The size separation of GQDs was confirmed by transmission electron microcopy images. PL excitation maps of chromatographically-separated GQDs show four distinct emission peaks at ~330, ~440, ~520, and ~600 nm, respectively. The dominant luminescence features of the separated GQDs show discrete change depending on the overall size of GQDs, indicating that PL variation occurs because of differences in density, shape, and size of $sp^2$ fragments available in the GQDs. On the basis of the experimental results of PL peak wavelength and pH dependence, the emission is attributed to quasi-molecular PL from the fragments composed of a few aromatic rings with oxygen containing functional groups.

**TEXT**

Graphene with an atomically-thin carbon honeycomb lattice structure has attracted a great deal of interest from the viewpoints of both fundamental studies and applications.[1-9] However, graphene is a zero-band-gap material, which limits optoelectronic applications in the visible light region.[10] A band gap can be introduced into graphene by the quantum confinement effect in the form of one-dimensional graphene nanoribbons and zero-dimensional graphene quantum dots (GQDs).[11-13] GQDs are nanometer-sized graphene structures and their band gap is expected to be controlled by changing the overall size as in the case of semiconductor QDs.[14] Various excellent properties of GQDs have been reported such as luminescence stability, high water solubility, and low toxicity, which make GQDs promising candidates for future applications such as in bioimaging,[15-17] photocatalysts,[18] and optoelectronics.[19,20]

Luminescence is one of the intriguing properties of GQDs.[15,21,22] Several synthesis methods for luminescent GQDs have been reported and can be classified into two groups: top-down[15,21,23] and bottom-up[19,24,25] routes. The large variation in size and chemical structure of fabricated GQDs depending on the synthetic process used makes it difficult to understand the mechanism of their luminescence. So far, several luminescence mechanisms of GQDs have been proposed to arise from the quantum size effect,[11,26] zigzag edge sites,[15,22] recombination of localized electron-hole pairs,[7,27] and the defect effect.[11,28] The size effect of GQDs is strongly expected to affect their optical properties, and understanding the relationship between the size and optical properties is essential for future optical applications. However, despite its importance, size separation and detailed size-dependent luminescence properties of GQDs have not been fully understood.

Here, we study the photoluminescence (PL) properties of GQDs separated by high performance liquid chromatography (HPLC). PL excitation (PLE) maps show that the PL of GQDs is composed of four distinct features, and PL properties of GQDs discretely change with changing the overall size. Discrete changes in emission wavelength and pH-dependent PL behavior indicate that the origins of PL features come from the small $sp^2$ fragments with oxygen containing functional groups embedded in the GQDs.

## RESULTS AND DISCUSSION

The GQDs used in this study were fabricated by a chemical oxidation process of pitch-based carbon fibers.[15] The thickness of as-prepared GQDs is ~1 nm, as measured by atomic force microscopy (AFM), indicating that most of the GQDs are of single-layer or bi-layer structure (Figure S1). Then, the as-prepared GQDs were separated by size-exclusion HPLC. Figure 1a shows a HPLC chromatogram of as-prepared GQDs monitored by optical absorption at 254 nm. The process of HPLC collection was started at the onset of the chromatogram (46 min) and each successive fraction was named fr 1–fr 15. Here, we mainly discuss the results of fr 4, fr 7, and fr 10 as typical samples indicated by the green, red, and blue areas in the figure, respectively. Figure 1b shows a photograph of a series of GQDs in solution separated by HPLC (HPLC-GQDs), where each fraction is displayed in order of its retention time (fr 1–fr 15). One can clearly see that the color of HPLC-GQDs changes depending on retention time.

The insets of Figures 2a and 2b show typical transmission electron microscope (TEM) images of GQDs in fr 4 and fr 9, which are observed as darker regions. The typical size of GQDs in fr 4 is larger than that in fr 9, as verified by the TEM observations. Figures 2a and 2b show histograms of the size distribution in the fr 4 and fr 9 GQDs observed in the TEM images. As expected by the principle of size-exclusion chromatography, both the TEM images and histograms suggest that the overall size of the GQDs decreases with increasing HPLC retention time. The average sizes of the GQDs within fr 4, fr 9, and the as-prepared GQDs are 10.8, 6.8, and 8.2 nm, respectively. The size distribution of as-prepared GQDs is broader relative to the separated GQDs, as shown in Figure S2. It is confirmed that the contrast observed in various TEM images comes from carbon nanostructures by the electron energy loss spectroscopy (EELS) C1s spectrum (Figure S3).

Figure 3a shows X-ray photoelectron spectroscopy (XPS) C1s spectra of the as-prepared GQDs and HPLC-GQDs. The analysis of the C1s spectrum of fr 10 is also shown in Figure 3a. The spectral shapes of C1s peaks among fractions (fr 4–fr 10) are similar, and the analysis of the peaks reveals that as-prepared GQDs and HPLC-GQDs contain peaks arising from C–C (285.0 eV), C–O (286.5 eV), C=O (287.3 eV), and COOH (288.6 eV) bonds. In addition, XPS spectra show that the intensities of the C1s and O1s peaks are almost unchanged among the fractions (Figure S4). Figure 3b shows the ratios of the functional groups of the GQDs as evaluated from the analysis of the C1s peaks as a function of the fraction number. The evaluated C/O ratios of HPLC-GQDs derived from the relative intensities of C1s and O1s peaks are also shown in the inset of Figure 3b. Both the functional groups and the C/O ratios are approximately constant among the fractions as a whole, indicating that the degree of oxidation and the concentrations of functional groups of the GQDs change little by HPLC separation.

Figure 4 shows absorption spectra of as-prepared GQDs and HPLC-GQDs. The absorbance of GQDs gradually increases with decreasing wavelength for each sample. The absorption spectra of HPLC-GQDs gradually change depending on the retention time (*i.e.*, with the size of the GQDs). A distinct energy gap and peak structure are not observed for the GQDs of fr 4, while peak structures are observed for the GQDs of fr 7 and fr 10 at ~300 nm, corresponding to n–π* transitions of nonbonding electrons in the C=O bonds.[29] These results show that GQDs with different absorption properties are separated through size-exclusion chromatography.

Figure 5a shows the PL intensity of as-prepared GQDs as a function of excitation and emission wavelength plotted as two-dimensional maps (PLE maps). The PLE map shows that PL of the as-prepared GQDs is composed of three distinct PL features. The emission wavelength of each feature is ~600, ~520, and ~440 nm, indicated as $P_A$, $P_B$, and $P_C$, respectively.

Figures 5b–d show the PLE maps of the GQDs of fr 4, fr 7, and fr 10. The two distinct peaks, $P_A$ and $P_B$, are observed in the near-infrared to visible emission wavelength region, which are dominant features of the PLE map for fr 4. The PLE map of the GQDs of fr 7 shows the four distinct emission peaks, $P_A$, $P_B$, $P_C$, and $P_D$ (where $P_D$ ~330 nm), wherein $P_B$ and $P_C$ are the dominant peaks. Peaks $P_C$ and $P_D$, mainly in the UV region, are dominant in the PLE map of fr 10. These experimental results clearly show that size-exclusion HPLC separates GQDs having distinct and different PL properties. A comparison among the three PLE maps reveals the presence of a blue shift in the dominant emission features with increasing HPLC retention time (*i.e*., decreasing overall size of GQDs). PLE behavior is different between peak $P_A$ and peaks $P_B$, $P_C$, and $P_D$ wherein the emission wavelength of $P_A$ depends on the excitation wavelength and that of peaks $P_B$, $P_C$ and $P_D$ is independent of the excitation wavelength. This result suggests that the origin of PL is different for $P_A$ than for peaks $P_B$, $P_C$, and $P_D$, as discussed below.

Figure 6a shows PL spectra of GQDs normalized by their maximum peak heights and obtained from excitation wavelengths of 280 (fr 10), 360 (fr 8), 480 (fr 7), and 560 nm (fr 7). Distinct PL peaks $P_A$ (~600 nm), $P_B$ (~520 nm), $P_C$ (~440 nm), and $P_D$ (~330 nm) are clearly observed in each spectrum, which covers the spectrum range from near-infrared to UV region. The inset of Figure 6a shows a photograph of PL of HPLC-GQDs for different excitation wavelengths and fractions, where blue, green, and yellow PL colors corresponding to $P_C$, $P_B$, and $P_A$ are clearly observed.

Figure 6b shows normalized PLE spectra of peaks $P_D$, $P_C$, $P_B$, and $P_A$, monitored at emission wavelengths of 330 (fr 10), 440 (fr 8), 520 (fr 7), and 600 nm (fr 7), respectively. The PLE spectra exhibit distinct peak structures and differ from the absorption spectra shown in Figure 3. The PLE spectra of $P_D$, $P_C$, $P_B$, and $P_A$ show several peaks as shown in Table 1. We also measured the decay time of each PL feature (Figure S5). Typical PL decay times are ~2–3 ns, and the decay time of each PL feature is also summarized in Table 1.

**Table 1.** PLE peaks and decay time of each PL feature.

| PL feature | PLE peaks (nm) | Decay time (ns) |
|---|---|---|
| $P_A$ | 265, 320, 560 | 2.8 |
| $P_B$ | 265, 320, 470 | 3.6 |
| $P_C$ | 265, 290, 370 | 1.7 |
| $P_D$ | 240, 280 | 3.3 |

Here, we discuss the origin of the PL features of the GQDs. According to the previous study using scanning tunneling microscopy and spectroscopy,[30] a band gap of 0.2 eV due to the quantum confinement effect is observed in nanometer-sized GQDs (5 nm) fabricated from carbon dots by a thermal annealing process under ultra-high vacuum conditions. In contrast, the GQDs studied here, which have a typical size (>~5 nm), show absorption and PL above 2 eV (~600 nm), which is much higher than the expected value of the band gap because of the quantum confinement effect. Hence, the PL properties of the GQDs used in this study, which are fabricated by oxidation under strong acidic conditions, are not directly determined by the overall size of the GQDs (>~5 nm).

The experimental fact of a PL energy above 2 eV suggests that the size of the luminescent fragments in the GQDs are much smaller than the overall size of GQDs (>~5 nm). As confirmed by the XPS measurements shown in Figure 2, the GQDs used in this study are composed of $sp^2$ regions surrounded by $sp^3$ oxidized regions because of the fabrication by strong acidic processing of carbon fibers. Calculations of the energy gap of π–π* transitions based on density functional theory[27] estimate the energy gap for a single benzene ring as ~7.4 eV (170 nm). The energy gap drastically decreases down to ~2.6 eV (480 nm) for five aromatic rings and exhibits discrete values for fewer than five rings. The energy gap changes moderately with further increasing number of aromatic rings. In the experimental observations, the wavelengths associated with the PL peaks ($P_A$, $P_B$, $P_C$, and $P_D$) show discrete values, indicating that the PL of GQDs originates from much smaller $sp^2$ fragments embedded in GQDs in comparison with overall size of the GQDs (>~5 nm).

$P_A$ could originate from sp$^2$ regions of various sizes and shapes larger than those regions from which the other PL features ($P_B$, $P_C$, and $P_D$) originate. Larger aromatic ring structures have greater variation in shapes and band gaps, which causes the excitation-dependent PL behavior and continuous change in the resonance excitation wavelength, as shown in Figure 5. On the other hand, $P_B$, $P_C$, and $P_D$ are molecular-like (quasi-molecular) PL features independent of the excitation wavelength.[31] These features are attributed to structures composed of a small number of benzene rings whose band gap shifts discretely with changing number of rings.

Here, we discuss the remarkable change of size-dependent PL features in the GQDs. The dominant PL features shift discretely according to the characteristics of the HPLC fractions, and this experimental result suggests that HPLC-GQDs with different sizes have different populations of sp$^2$ fragments available within the GQDs. Since the XPS analysis shown in Figure 2 indicates that the oxidization level and the ratio of oxygen containing functional groups are approximately constant among HPLC-GQDs, the discrete PL change must be caused by the population change of sp$^2$ carbon regions because of the size variation of the GQDs. The large GQDs possess a large probability of having larger sp$^2$ fragments. Hence, the PL feature $P_A$ from relatively larger aromatic ring structures is dominant for the fractions composed of large GQDs. On the contrary, the small GQDs do not have the overall size sufficient to form large sp$^2$ fragments. Thus, $P_C$ and $P_D$ derived from smaller sp$^2$ fragments are dominant in the smaller GQDs.

To further explore the origins of the PL observed, we studied the pH-dependence of the PL spectra. Figure S6 shows the relative integrated PL intensities of $P_A$ (fr 7), $P_B$ (fr 7), $P_C$ (fr 8), and $P_D$ (fr 10) excited at wavelengths of 560, 320, 280, and 280 nm, respectively, as a function of pH, where each point of the integrated PL intensity is normalized by absorbance at each excitation wavelength. Peaks $P_A$ and $P_B$ exhibit strong PL at a neutral condition and are quenched in acidic and basic conditions. In contrast, the PL intensities of peaks $P_C$ and $P_D$ become strong under basic conditions and weak under acidic conditions, which is a similar tendency to that previously reported.[22] These pH-dependent PL behaviors suggest that the origin of PL is small $sp^2$ fragments conjugating with oxygen containing functional groups such as carboxylic groups in the GQDs where the chemical structure of these functional groups depends on the pH of the solution.

**CONCLUSIONS**

In summary, we studied the PL properties of GQDs separated by the size-exclusion HPLC method. We demonstrated the controllability of optical properties through size separation of GQDs. The PLE maps of HPLC-GQDs clearly show that GQDs with different PL properties can be separated through HPLC. Discrete changes in the PL of HPLC-GQDs are demonstrated by changes in the relative intensities of the four distinct PL features corresponding to the emission wavelengths of ~600, ~520, ~440, and ~330 nm, respectively. The dominant luminescence features of the separated GQDs show discrete changes depending on the overall size of GQDs, indicating that PL variation occurs because of various densities, sizes, and shapes of $sp^2$ fragments available in the GQDs. The experimental results of PL peak wavelengths and pH dependence indicate that the emission originates from quasi-molecular PL derived from $sp^2$ fragments composed of small aromatic ring structures with oxygen containing functional groups. Our findings provide important insights for understanding the optical properties of GQDs synthesized by a strong oxidation process and open the door to new optical applications of GQDs.

**METHODS**

*Preparation of GQDs.* Pitch-based carbon fibers (0.15 g, purchased from *Fiber Glast Development Corporation*) were added into a mixture of concentrated $H_2SO_4$ (30 mL) and $HNO_3$ (10 mL).[15] The solution was sonicated for 2 h and stirred for 1 h at 100 °C. The mixture was cooled and diluted with distilled water (400 mL), then neutralized by $Na_2CO_3$.

*Chromatographic separation of GQDs.* Chromatographic separation was conducted according to the procedures described previously.[32] The following three columns were connected for HPLC separation of GQDs: Cosmosil CNT-2000, CNT-1000, and CNT-300 (diameter: 7.5 mm, length: 300 mm, Nacalai Tesque, Inc.) with pore sizes of 2000, 1000, and 300 Å, respectively. A phosphate buffer (20 mL, pH 7.0) containing $Na_2SO_4$ (100 mL) for optical measurement samples and milli-Q water for the TEM and XPS measurements were used as the mobile phase with a flow rate of 1.0 mLmin$^{-1}$. After injection of the GQD solution into the buffer (1.0 mL), elution was monitored by optical absorption at a wavelength of 254 nm. Collection of the fractions was conducted every minute for 15 min. Each fraction for TEM and XPS measurements was dialyzed in a dialysis bag (retained molecular weight: 1000 Da, Spectra/Por 7, Spectrum Laboratories, Inc.).

*Characterization of GQDs.* Absorption spectra were measured using a UV–Vis spectrophotometer (UV-1800, Shimadzu). The PL and PLE spectra were obtained using a fluorescence spectrophotometer (RF-5300PC, Shimadzu). The relative PL intensity was corrected for instrumental validations of detection sensitivity using standard lamps. The TEM images were obtained on a conventional electron microscope (JEM-2100F, JEOL) operated at 80 keV. The XPS measurements were conducted on X-ray photoelectron spectrometer (JPS9010-TRX, JEOL) with an Al cathode as the X-ray source. The samples were drop-casted on Si/SiO$_2$ substrates for XPS measurements. The height of GQDs was measured by AFM (NanoScope IIIa, Digital Instruments). Time-resolved PL spectroscopy was performed with a time-correlated single photon counting module (Hamamatsu Photonics, Quantaurus-Tau).

*Conflict of Interests:* The authors declare no competing financial interest.

*Supporting Information available:* An AFM image and corresponding height profile, a TEM image and corresponding size distribution histogram of as-prepared GQDs, an EELS spectrum, wide energy range XPS spectra, time-resolved PL decay curves, and the pH dependence of the PL intensity.

*Acknowledgment*: The authors thank E. Nakata and T. Morii for providing experimental equipment and N. Sasaki for XPS measurements. This study was supported by a Grant-in-Aid for Scientific Research from the Japan Society for the Promotion of Science (Grants No. 22740195, No. 22016007, No. 23340085, and No. 24681031), the Precursory Research for Embryonic Science and Technology program from the Japan Science and Technology Agency, The Asahi Glass Foundation, and the Yamada Science Foundation.

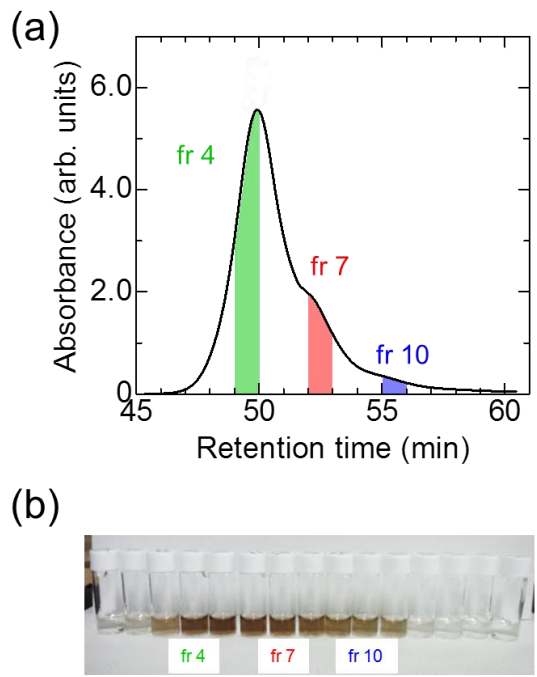

**Figure 1.** (a) HPLC chromatogram of GQDs monitored using optical absorption at a wavelength of 254 nm. The experimental results for three-typical fractions (fr 4, fr 7, and fr 10) indicated by green, red, and blue hatched regions are mainly discussed in this study. (b) A photograph of GQD solutions separated by HPLC (HPCL-GQDs) arranged in the order of retention time.

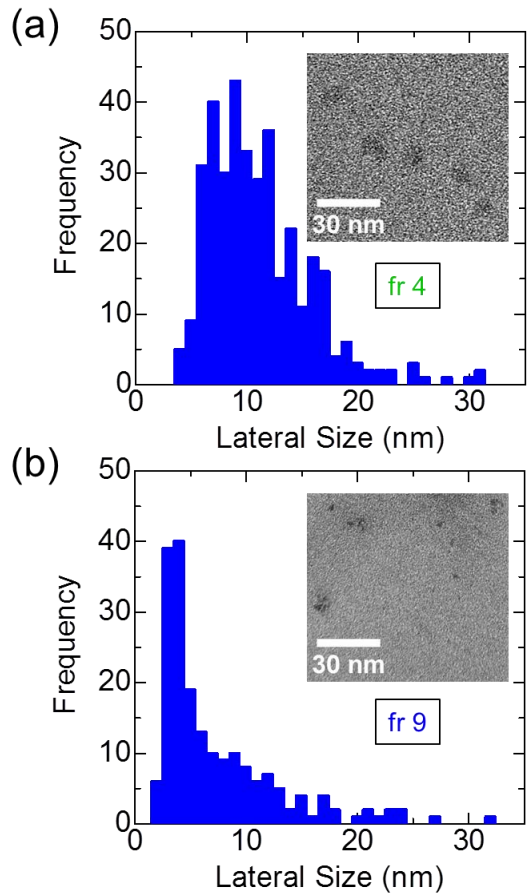

**Figure 2.** Size distribution histograms of (a) fr 4 and (b) fr 9 GQDs. Insets show the typical TEM images of GQDs in each fraction.

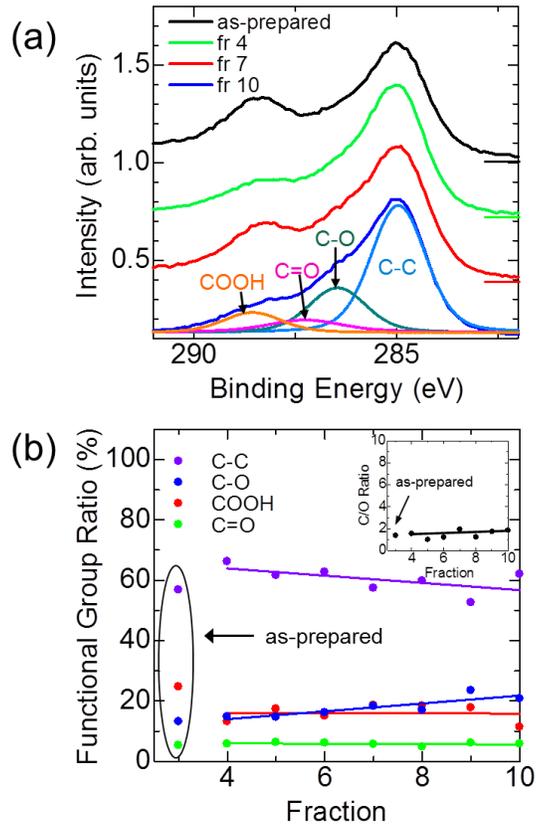

**Figure 3.** (a) XPS C1s spectra of as-prepared, fr 4, fr 7, and fr 10 GQDs. The analysis of the C1s spectrum of fr 10 is also shown. (b) The ratio of functional groups of as-prepared and HPLC-GQDs as a function of the fraction number. The inset shows the C/O ratio of as-prepared and HPLC-GQDs as a function of the fraction number.

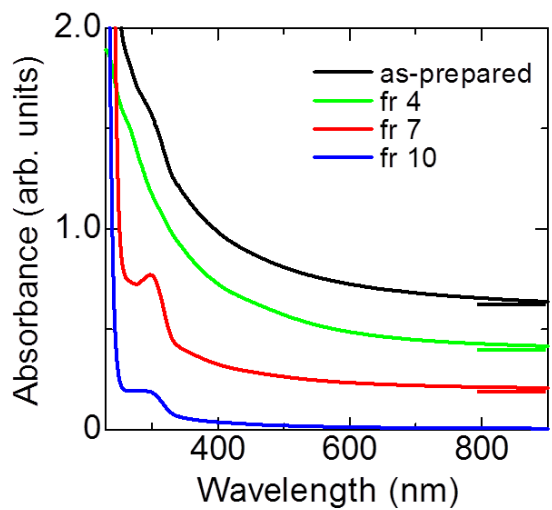

**Figure 4.** Absorption spectra of as-prepared (black line), fr 4 (green line), fr 7 (red line), and fr 10 (blue line) GQDs.

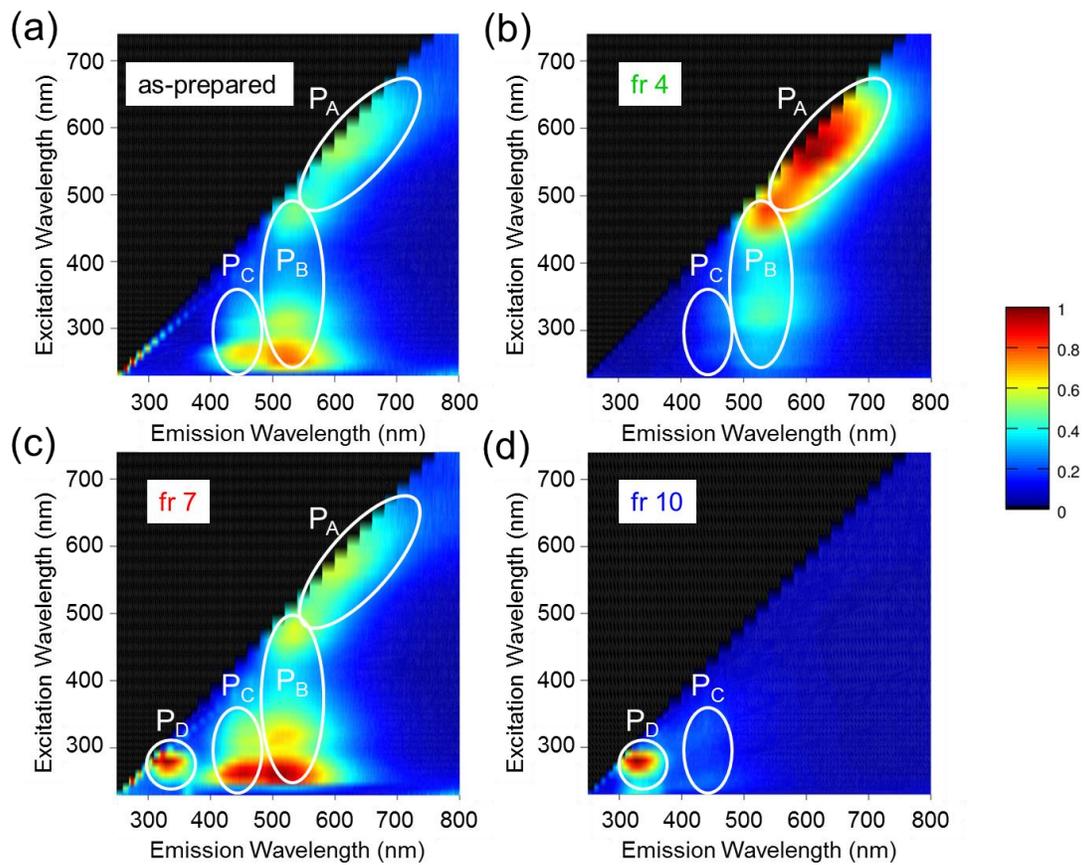

**Figure 5.** PLE maps of (a) as-prepared, (b) fr 4, (c) fr 7, and (d) fr 10 GQDs. Four PL features ($P_A$, $P_B$, $P_C$, and $P_D$) are indicated in each map by white ellipses.

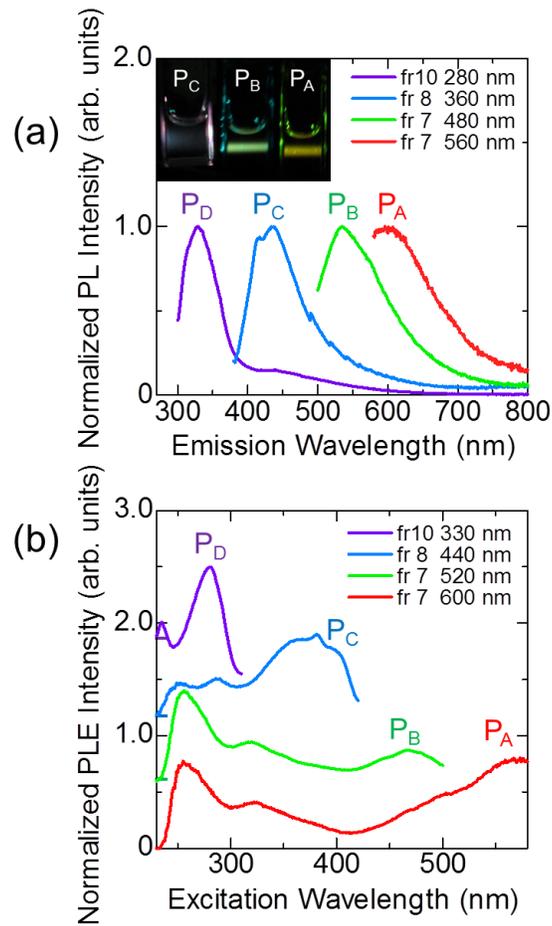

**Figure 6.** (a) Typical normalized PL spectra of HPLC-GQDs excited with wavelengths of 280 (fr 10), 360 (fr 8), 480 (fr 7), and 560 nm (fr 7). The inset shows a photograph of HPLC-GQDs exited by a $D_2$ lamp (fr 9) and at wavelengths of 480 and 560 nm (fr 4) by a super-continuum laser light source, corresponding to $P_C$, $P_B$, and $P_A$, respectively. (b) Normalized PLE spectra monitored at 600 ($P_A$, fr 7), 520 ($P_B$, fr 7), 440 ($P_C$, fr 8), and 330 nm ($P_D$, fr 10).

**Supporting Information for**

# Size-Dependent Luminescence Properties of Chromatographically-Separated Graphene Quantum Dots

Naoto Fuyuno,[†] Daichi Kozawa,[†] Yuhei Miyauchi,[†,‡] Shinichiro Mouri,[†] Ryo Kitaura,[§] Hisanori Shinohara,[§] Toku Yasuda,[∥,⊥] Naoki Komatsu,[∥] and Kazunari Matsuda[†,*]

[†]Institute of Advanced Energy, Kyoto University, Uji, Kyoto 611-0011, Japan

[‡]Japan Science and Technology Agency, PRESTO, 4-1-8 Honcho Kawaguchi, Saitama 332-0012, Japan

[§]Department of Chemistry and Institute for Advanced Research, Nagoya University, Nagoya 464-8602, Japan

[∥]Department of Chemistry, Shiga University of Medical Science, Otsu 520-2192, Japan

[⊥]Department of Materials Science, Osaka Electro-Communication University, Neyagawa, Osaka 572-8530, Japan

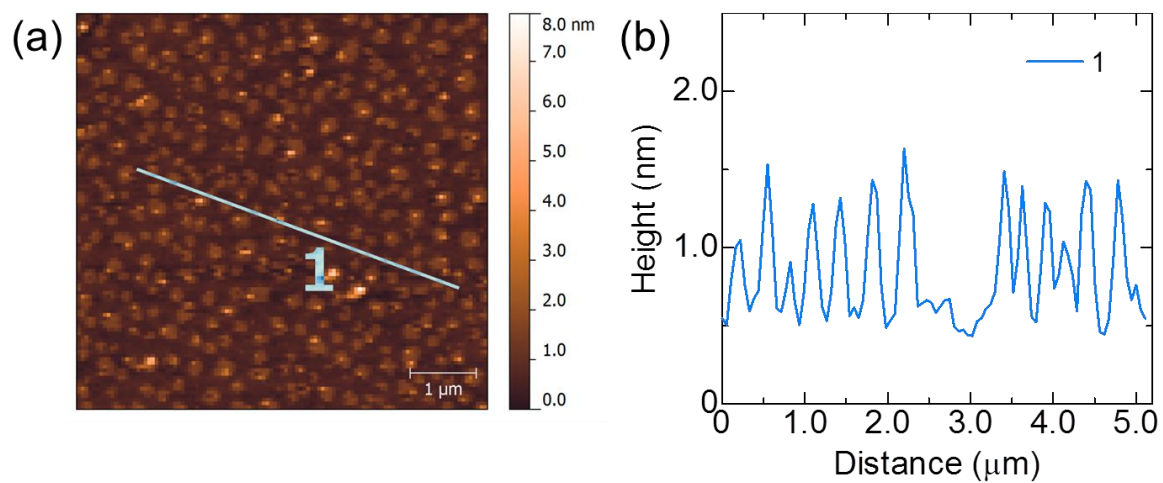

**Figure S1.** (a) An AFM image of as-prepared GQDs and (b) A height profile along line 1 shown in (a).

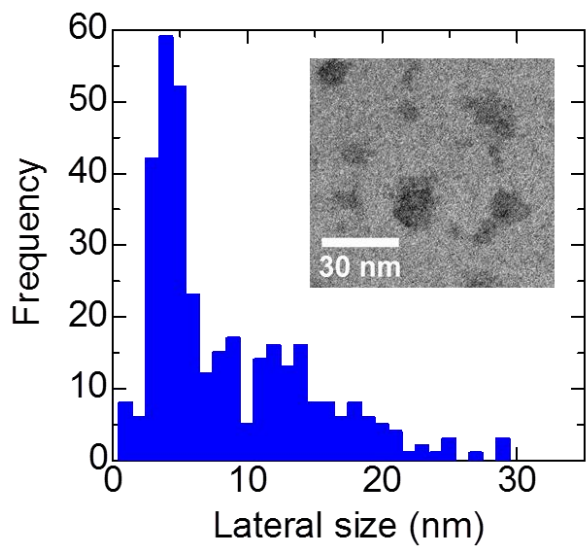

**Figure S2.** A size distribution histogram of as-prepared GQDs. The inset shows a TEM image of as-prepared GQDs.

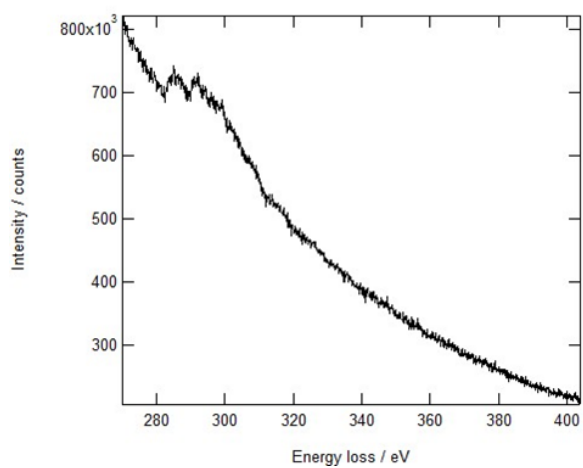

**Figure S3.** An EELS spectrum of as-prepared GQDs.

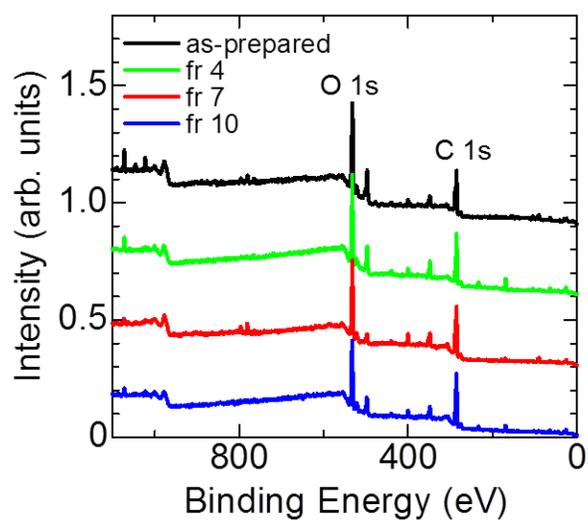

**Figure S4.** Wide energy range XPS spectra of as-prepared, fr 4, fr 7, and fr 10 GQDs.

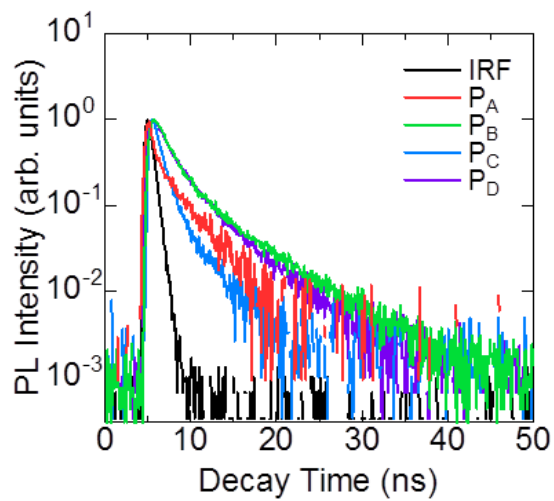

**Figure S5.** Time-resolved PL decay curves of $P_A$, $P_B$, $P_C$, and $P_D$ excited by wavelengths of 470, 280, 340, and 280 nm, respectively. The black line indicates the instrument response function (IRF).

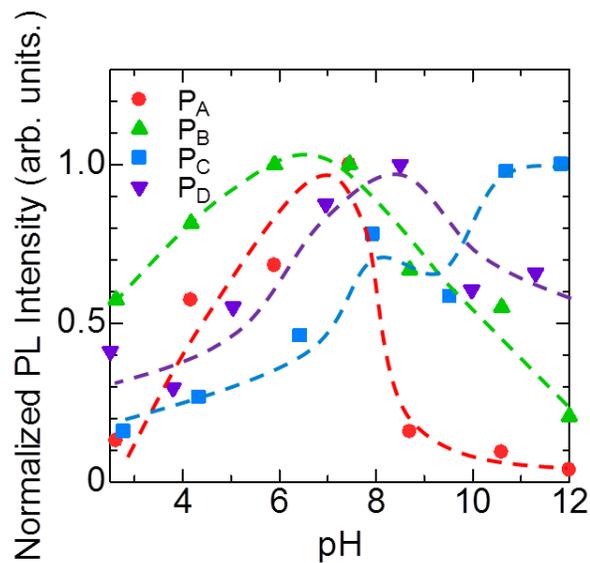

**Figure S6.** Relative integrated PL intensity of four-distinct peaks ($P_A$, $P_B$, $P_C$, and $P_D$) as a function of pH. Dotted curves are guides for the eyes.

The transient PL measurements were conducted to obtain PL lifetimes (Figure S5). The PL lifetimes of each decay curve are determined as an average lifetime $\tau_{PL}$. The PL lifetime can be written as

$$\tau_{PL} = \int_0^\infty tI(t)dt \bigg/ \int_0^\infty I(t)dt, \qquad (1)$$

where $t$ is the delay time and $I(t)$ is the PL intensity as a function of $t$. The PL lifetime of each feature in nanoseconds is coincident with a previous result.[1]